# Applying Data Privacy Techniques on Tabular Data in Uganda

Kato Mivule
*Computer Science Department*
*Bowie State University*
*14000 Jericho Park Road Bowie, MD 20715, USA*
mivulek0220@students.bowiestate.edu

Claude Turner
*Computer Science Department*
*Bowie State University*
*14000 Jericho Park Road Bowie, MD 20715,USA*
cturner@bowiestate.edu

## ABSTRACT

The growth of Information Technology(IT) in Africa has led to an increase in the utilization of communication networks for data transaction across the continent. A growing number of entities in the private sector, academia, and government, have deployed the Internet as a medium to transact in data, routinely posting statistical and non statistical data online and thereby making many in Africa increasingly dependent on the Internet for data transactions. In the country of Uganda, exponential growth in data transaction has presented a new challenge: What is the most efficient way to implement data privacy. This article discusses data privacy challenges faced by the country of Uganda and implementation of data privacy techniques for published tabular data. We make the case for data privacy, survey concepts of data privacy, and implementations that could be employed to provide data privacy in Uganda.

## 1. INTRODUCTION

The growth of Information Technology (IT) in Africa has led to an increase in data transaction across Africa's communication networks, with 110 million Internet users, 17 million Facebook accounts, and a penetration rate of 10 percent for Internet and 1.7 percent penetration rate for Facebook in Africa as of August 2010. Uganda alone boasts 3.2 million Internet users, 192,000 Facebook accounts with penetration rates at 9.6 and 0.6 percent for Internet and Facebook respectively. The growth rates are only expected to exponentially increase[International Telecommunications Union, 2009, 2010]. In Uganda's case, higher education institutions routinely post student admission and graduation data online and grant access to student records online [Makerere University Admission List, 2010]. The Ugandan Electoral Commission posted the National Voter's Register Online[The Electoral Commission of Uganda, 2010; U.S Embassy Kampala Uganda, 2010]. While the Uganda Bureau of Statistics publishes statistical data routinely, and takes great care to remove personal identifiable information (PII), a review of the published data sets from other Ugandan entities such as educational institutions and the Electoral Commission of Uganda, show PII was included in published data sets. At the same time a growing number of young Ugandans are fans of large Online Social Networks (OSN) like Facebook, resulting in large amounts of PII leaked from the online auxiliary data sources.

In developed countries like the USA, data gathering institutions are bounded by state and federal privacy laws that require that privacy of individuals be protected. One example in the USA is the Privacy Act of 1974, Health Insurance Portability and Accountability Act (HIPAA) of 1996, and Personal Data Privacy and Security Act of 2009, that would require entities to protect and secure PII in data [USDOJ, 1974; USGPO, 1996; US Library of Congress, 2009]. While Uganda's constitution defines the rights of an individual to privacy in terms of interference, stating that no person shall be subjected to interference with the privacy of that person's home, correspondence, communication or other property, no precise definition is given in the context of PII, data privacy, and computer security [Embassy of The Republic of Uganda, 2010; Parliament of Uganda, 2010].

Ugandan Bureau of Statistics Act of 1998 describes Ugandan government policy on data collected by the Ugandan Bureau of Statistics(UBS). Absent from that description is how none governmental entities collect and disseminate data. The Ugandan Bureau of Statistics Act of 1998 does not discuss what PII is in the Ugandan context. The only close reference is the "removal of identifiers" before data is granted to researchers [ Uganda Bureau of Statistics, 1998 ]. In this case "identifiers" is ambiguous and could perhaps reference 'names' but not 'geographical location'. However, UBS with expert care, does publish de-identified micro data sets online but at the same time, many entities in Uganda publish none de-identified tabular data sets. A look at documents from authorities that govern communication technology in Uganda, the

Uganda Communications Commission (UCC) and the Ministry of Information and Communications Technology (ICT) show that policies on Data Privacy and Security have not been clearly formulated [Uganda Communications Commission, 2010; Privacy International, 2007; Ministry of Information and Communications Technology, 2009; Ministry of Information and Communications Technology, 2006; Uganda Communications Commission, 2000; Ministry of Works Housing and Communications, 2003]. In the USA for instance, PII could include an individual's social security number yet in Uganda, social security numbers are non existent, thus the set of PII in the USA differs from that in Uganda. Therefore, there is a need to expand Uganda's policy on how government and non-government entities collect and disseminate data. To this date no clear legal and technological data privacy frameworks exist in Uganda, and the same can be attested to many sub-Saharan Africa nations, Uganda being a case study. Despite the absence of a clear formulated policy on data privacy, we suggest application of data privacy techniques that could be utilized in Uganda to provide basic data privacy.

## 2. BACKGROUND

### 2.1. Previous Work on Data Privacy in Uganda

There is little or no known literature on data privacy from Uganda and much of sub-Saharan Africa in general, given the relatively young and developing computing domain. At this time, to the best of our knowledge, we are the first to call for the application of data privacy techniques in Uganda. While research on computer security in Ugandan exists, most of the work centers on network accessibility control methodologies [Nakyeyune 2009; Mutebi and Rai 2010; Makori and Oenga 2010; Kizza et al 2010; Mirembe and Muyeba 2009]. While Mutyaba [2009] and Makori [2009] do an excellent presentation on cryptographic methodologies for computer security, and Okwangale and Ogao [2006] discuss data mining techniques, however, privacy preserving data mining (PPDM) methodologies are not covered. Bakibinga [2004] does an brilliant job at articulating the need for electronic privacy in Uganda from a policy view point. Frameworks for secure management of electronic records have been proposed by Luyombya [2010], Ssekibule and Mirembe [2007], and Kayondo [2009], however, the works focus on data security and access control. Yet still Data Privacy differs from from Data Security in that data privacy has to do with the confidentiality while data security focuses on the accessibility. Even when a database system is physically secured, an inference attack could occur on published data sets [Sweeney, 2002]. It should be noted that the Ugandan Bureau of Statistics Act of 1998 does provide a legal framework for data privacy that focuses on data gathered by the UBS. What is absent from the Ugandan computational literature is the data privacy technological framework that entities other than the Ugandan Bureau of Statistics, such as health, academia, and private business could employ[ Uganda Bureau of Statistics, 1998 ]. Therefore, it is in this light that we make the case for data privacy in Uganda and the need for more research on data privacy and PPDM methodologies tailored to the Ugandan and African context.

### 2.2. Essential data privacy terms

- *Data privacy* is the protection of an individual's or an entity's data against unauthorized disclosure [Ciriani et al, 2007 ]. *Data security* is the safety of data from illegitimate access, use, and alteration, focusing on access control while data privacy focuses on disclosure control [Denning and Denning, 1979].

- *Personally identifiable information* is any information about an individual that could be used to construct the full identity of that individual with or without their authorization; these are attributes that explicitly identify an individual. [U.S. Department of Homeland Security, 2008; Mccallister and Scarfone, 2010].

- *Data De-identification* is a process in which PII attributes are removed or transformed to such an extent that when the data is published, an individual's identity cannot be reconstructed [Ganta et al, 2008; Oganian and Domingo-ferrer, 2001].

- *Data utility verses privacy* has to do with how useful a published data set is to a consumer of that published data set, in other words data utility is a published data set that satisfies a user query [Rastogi et al, 2007; Sramka et al, 2010 ]. While data privacy is crucial, often the usefulness of data is lost when PII and none explicitly identifying attributes, also refereed to as quasi-attributes, are removed or transformed. Therefore a balance between privacy and data utility is always sought [Sankar, 2010]. It has been determined that achieving optimal data privacy while not distorting data utility is a continual challenge since all anonymization problems are considered NP-hard problems[Wong et al, 2007]. Organizations that collect large amounts of data often release data to the public in the form of s*tatistical databases*, that is, data that does not change, and in many cases the data is released in aggregated format [Adam and Wortmann, 1989].

- *Inference and reconstruction attacks* are methods of attack in which separate pieces of data are used to derive a

conclusion about a subject. The attacker will use a combination of publicly published data and other sources of separate information to reconstruct an individual's identity[Brewster, 1996].

- *Attributes*, in statistical databases, of which tabulated data is made up of, are field names or columns. Various types of attributes are taken into consideration when dealing with data privacy as pointed out by Ciriani et al. [ 2007]

- *PII attributes* are properties that uniquely identify an individual. Examples include SSN, and combination of first and last name. None explicitly identifying attributes or "q*uasi-attributes"* are attributes not in the PII category but can be used to reconstruct an individual's identity in conjunction with external data.

- *Confidential attributes* are attributes not in the PII and quasi-attributes category but contain sensitive information, such as disease, diagnosis, salary, HIV status, etc. *Non confidential attributes* are attributes that individuals or entities do not consider sensitive and whose publication does not cause disclosure. However, studies have shown that none confidential attributes can still be used to re-identify an individual given auxiliary data, thus making the explicit description of what PII is and is not even more challenging[Narayanan and Shmatikov, 2010].

## 2.3 Data privacy techniques

- *Data privacy methods* are categorized as *non-perturbative* techniques in which original data is not modified, some data is suppressed or some sensitive details removed while with p*ertubative techniques,* original data is altered or disguised so as to protect PII and sensitive data [Ciriani et al, 2007]. While a number of data privacy techniques exists, we focus on application of *k-anonymity*, *suppression*, and *generalization*.

- *Suppression is* a popular data privacy method in which data values that are unique and can be used to establish an individual's identity are omitted from the published data set [Bayardo and Agrawal, 2005; Ciriani et al, 2010]. However, even after some unique cell values have been suppressed, it is still possible for an attacker to use peripheral results such as averages and totals, to re-determine the value of the suppressed cell, thus the cell suppression problem(CSP), which is considered an NP-Hard problem. With the cell suppression problem, given a set of sensitive primary cells to be concealed, a set of additional secondary cells has to be found and suppressed so as to disallow an attacker from determining the values of the primary cells. [Castro, 2011; Matthews and Harel, 2011; Almeidaa et al, 2006]

- *Generalization* is a data privacy method in which attributes that could cause identity disclosure are made less informative; sensitive values are replaced with a general none revealing value. An example includes replacing the gender attribute value with "person" instead of "Male" or "Female" [Samarati and Sweeney, 1998]. Wang [2004] defines a valid generalization as being in the form of *{c}→p*, where all child values *{c}* are replaced with parent value *p*, as long as all values below *c* have been generalized to *c* [ Wang, 2004].

- Ciriani et al [2010] advance the description of generalization as a process of replacing the values of an attribute in a *PT[QI]* with unspecific and generic values, where each property *A* in *PT* is at the beginning connected with a ground domain *D=dom(A, PT)*. Where *QI* is the quasi-identifiers, *PT* is the private table. Each value $v \in D$ is represented with unspecific value $v' \in D'$, where *D'* is a non-specialized domain for *D;* the non-specialized domain for *D* is designated as *Dom*. Each domain $D \in Dom$, is a successive hierarchy known as the domain generalization hierarchy, symbolized by $DGH_D$, and diagrammatically portrayed by links of vertices [Ciriani et al, 2010].

- *k-anonymity* is a data privacy enhancing mechanism that utilizes *generalization*, and *suppression* as outlined extensively by Samarati [2001] and Sweeney [2002]. *k-anonymity* requires that for a data set with quasi-identifier attributes in database to be published, values in the quasi-identifier attributes be repeated at least *k* times to ensure privacy; that is, *k >1* [Sweeney, 2002]. However, achieving the optimal *k-anonymized* data set has been shown to be an NP-Hard problem [Meyerson and Williams, 2004].

- *l-diversity* first disclosed by Machanavajjhala et al, is an extension of *k-anonymity* that seeks to prevent information leak attacks on a homogenous attributes. For example, if a database contains a diagnosis status attribute, with each individual listed as HIV positive, then the attacker knows for sure that anyone registered in that database is HIV positive. To prevent this type of attack, Machanavajjhala et al proposed *l-diversity,* an improvement of *k-anonymity* that seeks to increase diversity among sensitive attributes by stating that a *q-block* is

*l-diverse* if it incorporates at most *l* "well-represented" values for the sensitive attribute **S**. A table is *l-diverse* if every *q*-**block** is *l-diverse*.[Machanavajjhala et al, 2006] *l-diversity* is mitigated if for each group of records sharing a combination of key attributes, there are at most *l* "well represented" values for each private attribute; however, researchers have found that *l-diversity* still suffers from skewness and similarity attacks [Domingo-ferrer and Torra, 2008].

- *Noise addition,* another perturbation technique, is a data privacy procedure defined by Kim [1986], in which a random numeric value (noise) is added to confidential numeric data values to provide concealment [Domingo-Ferrer et al, 2004; Islam, 2007; Kadampur and Somayajulu, 2010]. For example, we could conceal a student's 3.50 GPA by adding a random value of 0.25, thus publishing a 3.75 GPA[Huang et al, 2005]. Noise perturbation techniques have found applications in the creation of *synthetic data*, a process of producing data with all statistical similarities of the original data [ Mateo-sanz et al, 2004 ].

- *Differential Privacy* another perturbation technique that has recently gained attention in data privacy research, is a process in which Laplace noise is added to a query response such that the presence or absence of an individual cannot be observed. [Dwork 2011 ].

- *K-anonymity and Differential Privacy Hybrid*, is another recent technique that seeks to build on the weakness and strength of *k-anonymity* and *Differential Privacy* algorithms in regards to the NP-hard problem of utility verses privacy[Li et al, 2010]. However, in this article, we focus on k-anonymity, suppression, and generalization.

## 3. DATA PRIVACY EXPERIMENT

In this section we applied a combination of *k-anonymity, suppression, and generalization* techniques on a Ugandan data set of about 1200 records, to implement data privacy. In our lab experiment, we employed the *k-anonymity* methodology to de-identify data from a Makerere University admission tabular data published publicly by the University as Student Admission Records and posted online [Makerere University Admission List, 2010]. Our reason for choosing *k-anonymity* is the ease of implementation for tabular data privacy. Our concept is that academia, businesses, non governmental organizations with no highly skilled computational experts in Uganda, could easily implement *k-anonymity* to provide basic data privacy for tabular data; tabular data being the most common type of published data in Uganda.

### 3.1. Steps taken to implement Data privacy

Our initial step was to de-identify the data set by removing PII as defined by the US data privacy laws. While no explicit data privacy laws exist in Uganda, we utilized the definitions of what constitutes PII as defined by the US data privacy laws (HIPAA), considering that they are universally acceptable. After we removed PII, we identified attribute values that we could suppress and other attribute values we could generalize. We then applied *k-anonymity* to the de-identified data set and rechecked if there was need reapply *suppression and generalization* to satisfy *k-anonymity*. We then output the de-identified tabular set satisfying *k-anonymity*. We checked for data utility to see if data to be published is meaningful to the user while not compromising privacy [Rastogi et al, 2007; Sramka et al, 2010 ]. The procedure we employed is outlined next, and is illustrated in Figure 1.

PROCESS TAKEN TO ACHIEVE DATA PRIVACY
INPUT: Data from relation or schema
OUTPUT: Data privacy preserving published tabular data set
STEP 1. Identify PII Attributes
STEP 2. Remove PII Attributes
STEP 3. Identify none explicitly identifying or quasi-identifier attributes
STEP 4. Generalize or Suppress quasi-identifier attributes
STEP 5. Sort or order data
STEP 6. Check that *k>1* in tuples
STEP 7. Check for single values in attributes that cannot be grouped together to
    achieve *k>1*
STEP 8. If single values and outliers that cannot be grouped together still exist
    in attributes, then continue to Generalize or Suppress quasi-identifier attribute values
    until k-anonymity is achieved at *k>1*
STEP 9. Check for utility
STEP 10. Publish tabular data set

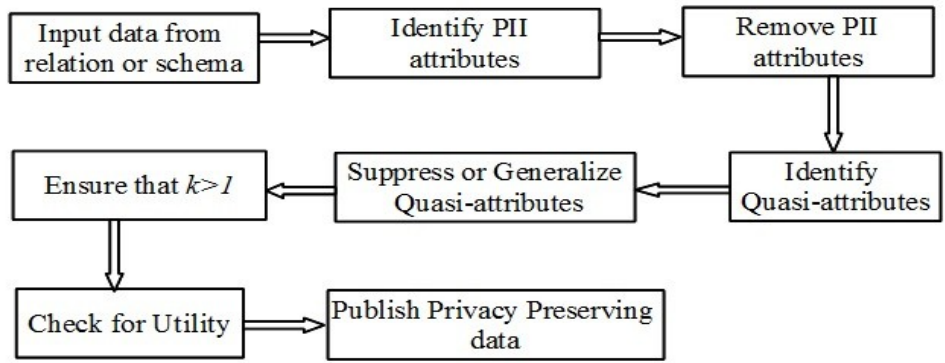

*Figure 1. A data privacy process utilizing k-anonymity*

The original published data set included the following attributes, in which we let:
- *A* = { RegNo, StudentNo, Lname, Fname, Mname, Sex, BirthDate, Nationality, Hall, Program, IndexNo, Year }, the relation *admission list* that included all attributes in the published data set.
- *B* = { Lname, Fname, Mname, StudentNo, IndexNo, RegNo}, the set of all PII attributes that we identified in the published data set.
- *C* = { Nationality, Sex, BirthDate,}, the set of all quasi-identifier attributes identified in the data set.
- *D*={Hall, Program, Year}, the set of all non-sensitive attributes.
- *E*={ }, the set of all sensitive attributes.
- Thus, we have: *B⊂ A, C⊂ A, D⊂ A* and *E⊂ A*
- Therefore *A=B∪ C∪ D∪ E, A ={ B, C, D, E}.*
- Removing PII yields *A ={ C, D, E}.*
- The de-identification of the *Admission List* set involves a complement of the *PII* set: $(B)^c = U – B = A – B = C + D + E.$
- Thus, therefore we remained with the *Quasi attributes, Non-Sensitive attributes, and Sensitive Attributes;* where *U* is the universal set, which in this case is all the *Admission List attributes.*
- We suppressed or generalized the *Quasi Attributes:* suppress or generalize (*C*);
- Then, we applied *k-anonymity*: k-anonymity( $(B)^c$ );
- **Finally,** we ordered values of $(B)^c$; If *k = 1*, we suppressed or generalized *C* until *k >1*.

### 3.2. Data utility challenges of removing PII
With the Makerere University data set, removing names and student numbers entirely kills utility. The data becomes meaningless to students who simply want to view it to see if their names are on the university admission list. One way this problem can be dealt with, is by publishing a list with just the *student number* or *student names* while obscuring other data as illustrated in the following two scenario:
- Scenario 1: we include *student number* publication of the university admission list:*Admission List = {StudentNo, Hall, Program, Year}.*
- Scenario 2: we include *student names* for publication of the university admission list: *Admission List = {Fname, Lname, Hall, Program, Year}.*

In both scenarios, the issue of balancing data utility and data privacy remain quite challenging and demand tradeoffs.

### 3.3. Relational model view
For a formal relational model view,
- we let $\pi$ *<attribute list>*$^{(R)}$
- where $\pi$ is the projection or selecting of attributes from a relation (Table),
- *<attribute list>* is the list of attributes from *Admission List,*
- $^{(R)}$ is the relation from which we select attributes.

- The original projection with all attributes is:
  - $\pi <RegNo, StudentNo, Lname, Fname, Mname, Sex, BirthDate, Nationality, Hall, Program, IndexNo, Year>^{(Admission\ List)}$.

The projection void of PII attributes is:
- $To\_Be\_Published\_List \leftarrow \pi< Sex, BirthDate, Nationality, Hall, Program, Year >^{(Admission\ List)}$.
- We applied k-anonymity to the list that is to be published:
  - k-anonymity(To_Be_Published_List).

| RegNo | StudentNo | Lname | Fname | Mname | Sex | BirthDate | Nationality | Hall | Program | IndexNo | Year |
|---|---|---|---|---|---|---|---|---|---|---|---|
| 09/U//EVE | 20900 | Annet | Anna | | F | 01/01/67 | UGANDAN | AFRICA | LIS | U0166 | 2008 |
| 09/U//EVE | 20901 | Green | RICE | | F | 01/01/80 | UGANDAN | MARY STUART | ARM | U0763 | 2008 |
| 09/U//EVE | 20902 | Timothy | NICE | | F | 01/01/81 | KENYAN | MARY STUART | BLE | U0063 | 2007 |
| 09/U//EVE | 20903 | Jones | Jane | GRACE | F | 01/01/73 | TANZANIA | MARY STUART | LIS | U0198 | 2007 |
| 09/U//EVE | 20904 | Carter | James | | M | 01/01/74 | UGANDAN | | RAM | U0160 | 2007 |
| 09/U//EVE | 20905 | Brown | Britain | N | F | 01/01/83 | KENYAN | AFRICA | ARM | U0715 | 2008 |
| 09/U//EVE | 20906 | Sams | Sam | | F | 01/01/84 | TANZANIA | MARY STUART | RAM | U0725 | 2007 |
| 09/U//EVE | 20907 | Faster | Master | | M | 01/01/85 | UGANDAN | | BLE | U1148 | 2008 |
| 09/U//EVE | 20908 | Uhuru | Kenya | | F | 01/01/90 | UGANDAN | COMPLEX | ARM | U0062 | 2007 |
| 09/U//EVE | 20909 | Vineyard | Martha | | M | 01/01/88 | KENYAN | AFRICA | ARM | U1017 | 2008 |

**Table 1.** *Admission List with PII – Data is fictitious for illustrative purposes*

### 3.4. The SQL Implementation
We implemented de-identification in SQL by creating a SQL View and doing SELECT on the view by choosing only attributes that remain in the *Admission List* after removing PII. We created SQL Views that are void of PII attributes:

CREATE VIEW V2 AS SELECT  Sex, BirthDate, Nationality, Hall, Program, Year FROM Admission_List;

### 3.4.1. Generalization
We generalized the *BirthDate* attribute to further prevent any reconstruction attacks by first developing a domain generalization hierarchy(DGH), as shown below, after which we implemented the generalization in SQL. We choose the DGH based on the oldest person in the data set, and built our *DGH to $B_4 = \{196*\}$*, giving protection for the individuals born in 1967.

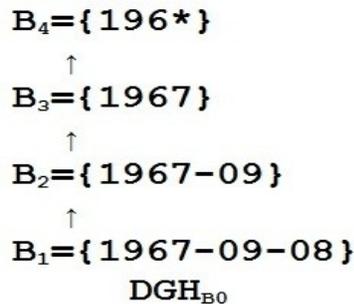

**Figure 2.** *Domain Generalization Hierarchy for the BirthDate Attribute.*

*SQL Implementation*
    CREATE table V2_Generalize1 SELECT Sex, BirthDate, Nationality, Hall, Program, Year FROM V2;
    UPDATE V2_Generalize1 set BirthDate ='196*' WHERE BirthDate BETWEEN 1967-01-01 AND 1999-12-31';

| Sex | BirthDate | Nationality | Hall | Program | Year |
|-----|-----------|-------------|-------------|---------|------|
| F | 196* | UGANDAN | AFRICA | LIS | 2008 |
| F | 196* | UGANDAN | MARY STUART | ARM | 2008 |
| F | 196* | KENYAN | MARY STUART | BLE | 2007 |
| F | 196* | UGANDAN | MARY STUART | LIS | 2008 |
| M | 196* | UGANDAN | | RAM | 2007 |
| F | 196* | KENYAN | AFRICA | ARM | 2008 |
| F | 196* | TANZANIA | MARY STUART | RAM | 2007 |
| M | 196* | UGANDAN | | BLE | 2008 |
| F | 196* | UGANDAN | COMPLEX | ARM | 2007 |
| M | 196* | TANZANIA | AFRICA | ARM | 2008 |

*Table 2. PII Attributes removed, BirthDate Attribute generalized to DGH to $B_4 = \{196*\}$*

### 3.4.2. Suppression
In the case of achieving *k-anonymity*, we had to suppress some values that appear once, yet still we had to ensure the utility of the data set.

| Sex | BirthDate | Nationality | Hall | Program | Year |
|-----|-----------|-------------|-------------|---------|------|
| F | 196* | UGANDAN | AFRICA | LIS | 2008 |
| F | 196* | UGANDAN | MARY STUART | ARM | 2008 |
| F | 196* | KENYAN | MARY STUART | BLE | 2007 |
| F | 196* | TANZANIA | MARY STUART | LIS | 2008 |
| M | 196* | UGANDAN | | RAM | 2007 |
| F | 196* | KENYAN | AFRICA | ARM | 2008 |
| F | 196* | TANZANIA | MARY STUART | RAM | 2007 |
| M | 196* | UGANDAN | | BLE | 2008 |
| F | 196* | UGANDAN | COMPLEX | ARM | 2007 |
| M | 196* | KENYAN | AFRICA | ARM | 2008 |

*Table 3. Highlighted values to be suppressed*

| Sex | BirthDate | Nationality | Hall | Program | Year |
|-----|-----------|-------------|-------------|---------|------|
| F | 196* | UGANDAN | AFRICA | LIS | 2008 |
| F | 196* | UGANDAN | MARY STUART | ARM | 2008 |
| F | 196* | KENYAN | MARY STUART | BLE | |
| F | 196* | TANZANIA | MARY STUART | LIS | |
| M | 196* | UGANDAN | | RAM | 2007 |
| F | 196* | KENYAN | AFRICA | ARM | 2008 |
| F | 196* | TANZANIA | MARY STUART | RAM | |
| M | 196* | UGANDAN | | BLE | 2008 |
| F | 196* | UGANDAN | | ARM | 2007 |
| M | 196* | KENYAN | AFRICA | ARM | 2008 |

*Table 4. We achieve k-anonymity at k>1*

    SQL: UPDATE V2_Generalize1 set Hall =' 'WHERE Hall ='Complex';

In **Table 3**, *k!>1* for Hall attribute. We suppressed the value 'Complex' in the Hall attribute, to achieve *k-anonymity* at **k>1** for all values in the attributes. Yet still even though the Year attribute satisfies *l-diversity*, still an attacker could single out a single record of a female from Kenya, a resident of Mary Stuart Hall, enrolled in 2007. Therefore, we employed suppression to further conceal such records while achieving *k-anonymity > 1* as illustrated in **Table 4**.

Check for *k-anonymity* that *k >1* by ordering data and counting that attribute values satisfy condition *k>1*:
    SELECT Sex, BirthDate, Nationality, Hall, Program, Year FROM V2 ORDER BY Sex, Program, Hall;

## 4. CONCLUSION
In this article we have discussed the concepts of data privacy, challenges faced by the country of Uganda when it comes to data confidentiality, and have suggested an implementation of data privacy for tabular data sets in Uganda. We have reviewed the state of data privacy policy in that nation state and some of the prior works done in the area of data privacy and security. Further, we have applied the *k-anonymity* data privacy methodology to Ugandan data sets. We have made the case for the need to revamp Uganda's data privacy policy to encompass both private and government sectors on how to gather and disseminate data, and the need to implement data de-identification techniques. With the growth of data transaction in Uganda, there is a need for more research on how to implement privacy preserving data publishing and privacy preserving data mining methodologies tailored to the Ugandan context, with applications ranging from academia, government, health sector, and private sector. While achieving optimal privacy yet maximizing utility continues to be an NP-hard problem, more studies need to be done on various implementations of optimal data privacy tailored to Ugandan context; with consideration that PII differs in Uganda from other geographical locations.